# On the Directivity of Low-Frequency Type IV Radio Bursts*

N. Gopalswamy, S. Akiyama, P. Mäkelä, S. Yashiro
Solar Physics Laboratory, Heliophysics Division
NASA Goddard Space Flight Center
Greenbelt, Maryland, USA
nat.gopalswamy@nasa.gov

I. H. Cairns
School of Physics
University of Sydney
Sydney, Australia
iver.cairns@sydney.edu.au

*Abstract*—An intense type IV radio burst was observed by the STEREO Behind (STB) spacecraft located about 144° behind Earth. The burst was associated with a large solar eruption that occurred on the backside of the Sun (N05E151) close to the disk center in the STB view. The eruption was also observed by the STEREO Ahead (STA) spacecraft (located at 149° ahead of Earth) as an eruption close to the west limb (N05W60) in that view. The type IV burst was complete in STB observations in that the envelope reached the lowest frequency and then receded to higher frequencies. The burst was partial viewed from STA, revealing only the edge coming down to the lowest frequency. The type IV burst was not observed at all near Earth because the source was 61° behind the east limb. The eruption was associated with a low-frequency type II burst observed in all three views, although it was not very intense. Solar energetic particles were also observed at both STEREOs and at SOHO, suggesting that the shock was much extended, consistent with the very high speed of the CME (~2048 km/s). These observations suggest that the type IV emission is directed along a narrow cone above the flare site. We confirm this result statistically using the type IV bursts of solar cycle 23.

## I. Introduction

Low-frequency (<14 MHz) type IV bursts are extensions of their metric counterparts and are relatively rare compared to type II and type III bursts in this frequency range. They typically extend down to a frequency of about 3 MHz. Observations in the frequency range 2-14 MHz became routinely available only in 1994 when the Wind spacecraft was launched with the Radio and Plasma Wave (WAVES) experiment [1] on board. For this reason, there have only been a few reports on the low-frequency type IV bursts [2-6]. One of the important characteristics of the type IV bursts is that they are associated mostly (~78%) with full halo coronal mass ejections (CMEs) [5]. It was not clear what aspect of halo CMEs is critical for the occurrence of type IV bursts. Front side full halo CMEs generally originate close to the disk center. Observing a type IV burst on 2013 November 7 in association with a CME observed by the inner (COR1) and outer (COR2) coronagraphs of the Solar Terrestrial Relations Observatory (STEREO [7]) and the Large Angle and Spectrometric Coronagraphs (LASCO [8]) on board the Solar and Heliospheric Observatory (SOHO) provides an important clue. The different appearance of the type IV burst when viewed from the two STEREO spacecraft and Wind suggest that the type IV emission is beamed along a narrow cone. In this paper, we provide detailed evidence that supports this suggestion.

## II. Observations

### A. Radio Observations

Figure 1 shows the radio dynamic spectrum from the WAVES instruments on board Wind and STEREO spacecraft. Wind was at Sun-Earth L1, while STEREO Ahead (STA) and Behind (STB) were located in Earth orbit at W149 and E144, respectively. The STB dynamic spectrum shows (i) a complex type III burst, (ii) a fragmented type II burst, (iii) a type IV burst between 16 and 8 MHz, and (iv) an enhancement of the type II burst between 18 and 21 UT. All these features are also observed by STA, except that only the initial edge of the type IV burst was observed. In the Wind/WAVES spectrum, the type IV burst was completely missing, while the type III bursts, the initial type II and its 18-21 UT extension were clearly observed. The type IV burst was observed as the emission at the top of the STB dynamic spectrum during 10:40 to 11:55 UT. The STB type IV burst is typical of these bursts reported previously [5] except that the simultaneous stereoscopic observations provide additional insights into the nature and origin of the burst.

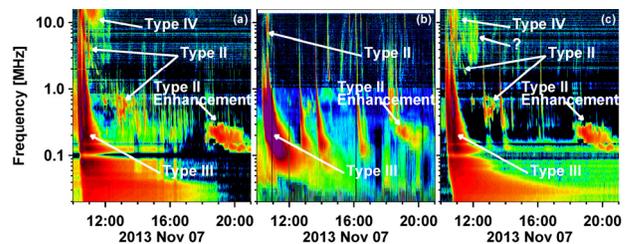

Fig.1. Radio dynamic spectra from STB (a), Wind (b), and STA (c) showing various bursts. The type IV burst is clear in (a), but not in (b). In (c), only a fraction of the type IV is observed. The type II is fragmented, but observed in all views.

*Work supported by NASA's Living with a Star TR&T program.

## B. White-Light and EUV Observations

The set of radio bursts in Fig.1 were associated with a fast and wide CME observed in white light originating on the backside of the Sun. The source was close to the disk center in STB view (N05E07), and near the west limb in STA view (N05W60). The source was behind the east limb (N05E151) in Earth (SOHO) view. STEREO's Extreme Ultra Violet Imager (EUVI) observed small changes in the source region as early as 9:25 UT. A CME appeared above the west limb in STA/EUVI image at 10:15 UT with the leading edge at a height of ~1.5 solar radii (Rs) and followed by a fast eruptive filament. In the STA/COR1 field of view (FOV), the CME first appeared at 10:25 UT at a height of ~2 Rs. Since the CME originated close to limb in STA view, we used observations from this spacecraft to measure CME kinematics. The CME was very fast (average speed ~2048 km/s in the combined FOV of STA's COR 1 and COR2).

Figure 2 shows the STEREO/COR2 and SOHO/LASCO snapshots of the CME in question. In STB/COR2 FOV, the CME was a full halo as it was heading towards that spacecraft. In STA, it was a wide, west-limb CME. In SOHO/LASCO, it was a wide, behind-the-east-limb CME, eventually becoming a full halo. The CME-driven shock arrived at STB around 13:30 UT on the next day, indicating a transit time of 27 h and a transit speed of ~1543 km/s. Given the in-situ shock speed at STB as ~860 km/s, the initial CME speed is expected to be ~2226 km/s suggesting deceleration from an initial high speed, consistent with the coronagraph observations. Fig. 3 shows the post-eruption arcade of the event extended in the east-west direction observed by STEREO/EUVI indicating the source location of the eruption. The longitudinal extension of the arcade was ~16°, typical of most large eruptions.

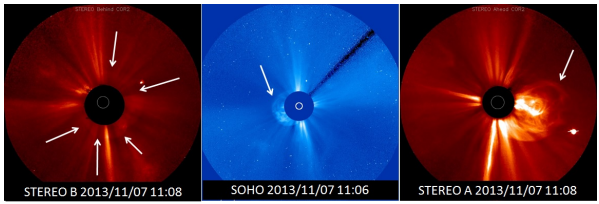

Fig. 2. Coronagraph images showing the CME as observed by STB (left), SOHO (middle), and STA (right). The arrows point to the CME. The inner white circles mark the optical size of the Sun. The dark disks are the images of the occulters in the coronagraph. The brightest feature in the right image corresponds to the eruptive prominence that became the core of the CME and followed closely behind the CME leading edge.

## III. RESULTS

### A. Directivity of the Type IV Burst

The eruption was a disk-center event in STB and a limb event in STA. STA and STB were separated by ~67° at the time of the eruption. The type IV burst is observed as a full event in STB, but as a partial one in STA. In Earth view, the type IV burst was absent, mainly because the source was occulted by the limb (the source was ~61° behind the east limb). All these observations point to a narrow cone of type IV burst associated with loops overlying the EUV post-eruption arcades of the eruption. As suggested in [5], the radio emission should originate from a heliocentric distance of ~3.5 to 4.5 Rs, depending on whether the radio emission occurs at the fundamental or harmonic of the plasma frequency. The 14 MHz plasma level (the highest frequency used by the Wind/WAVES experiment) occurs typically at a heliocentric distance of ~2 Rs. A source at this height would be occulted if it is located at ~60° behind the limb, as confirmed by the absence of the type IV burst in the Wind/WAVES dynamic spectrum. The burst was directed toward STB, so the full burst was observed. Viewed from STA, the foreground plasma seems to have absorbed all but the emission from the tallest loops. This interpretation is consistent with the small extent of the post-eruption arcades. The CME leading edge and prominence core were too far away when the type IV reached its lowest frequency (see Fig.2, STA/COR2 image), suggesting that the burst was not a moving type IV. Moving type IV bursts originate from CME structures that trap electrons accelerated at the flare site.

### B. The Type II Burst

The type II burst was fragmented, but the initial segment (from the highest frequency to 4 MHz) was observed in all three views. Unlike the post-eruption arcade loops, the CME-driven shock was much extended, and the emission was not occulted significantly by the shock. Furthermore, the eruption was associated with a large SEP event in STA and STB and a minor event at Earth, confirming the large extent of the shock. The type II burst enhancement observed around 18 UT in all three views (Fig. 1) suggests that the shock must have passed through an extended region with properties conducive to increased radio emission (possibly material from a preceding CME).

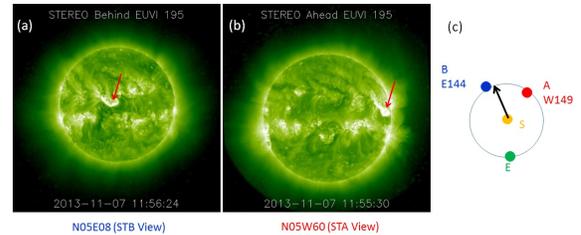

Fig. 3. EUVI images showing the post-eruption arcades (pointed by arrow) from STB (a) and STA (b) about an hour and a half after the CME launch. The circle in (c) is the Earth orbit with Earth (E), STA (A), and STB (B) marked. The black arrow indicates the direction of the CME from the Sun (S) heading toward STB. The solar disk in the STB image is slightly smaller because the spacecraft is slightly father from the Sun compared to STA.

### C. Type IV Bursts from Solar Cycle 23

In order to confirm the finding that type IV bursts from a disk-center source appear complete, while those from limb sources appear partial, we considered all the cycle-23 type IV bursts reported in [5]. We dropped a few events that were not clearly type IV bursts. In two more cases there was a CME data gap. We used the remaining 38 events for further analysis. Examples of disk-center and limb type IV bursts from cycle 23 are shown in Fig. 4a and 4b, respectively. The type IV burst in Fig. 4a occurred on 2005 January 15 around

8:30 UT. The burst was associated with an M8.6 flare and a fast (~2050 km/s) full halo CME originating from N16E04 (close to the disk center). The 2005 September 9 burst in Fig. 4b was associated with an X6.7 flare and also a fast (~2260 km/s) full halo CME from S12E67 (close to the east limb).

The solar sources of the 38 type IV bursts are shown in Fig. 4c. The size of the symbols indicates the size of the associated soft X-ray flares reported by GOES. Twenty bursts (or 53%) were associated with X-class flares, 14 with M-class flares (37%), and only 4 were associated with C-class flares (10%). The large number of M and X class flares is consistent with the high speeds of the CMEs. It is well known that the soft X-ray flare fluence and CME kinetic energy are moderately correlated [9].

We find that only 5 of the 38 events (13%) originated from close to the limb (shown by red symbols in Fig. 4c). In the non-limb cases, the type IV bursts appeared complete (similar to the example shown in Fig. 4a or the burst in STB view in Fig. 1). In the limb cases, the type IV bursts were partial (similar to the burst in Fig. 4b or the one in STA view in Fig. 1). Thirty of the 38 eruptions in Fig. 4c (or 79%) were halo CMEs. The large halo fraction is consistent with the source locations being close to the disk center, which seems to be important to observe the complete burst. Halo CMEs are more energetic, which is also confirmed by the fact that most of the associated flares were of X- and M-class.

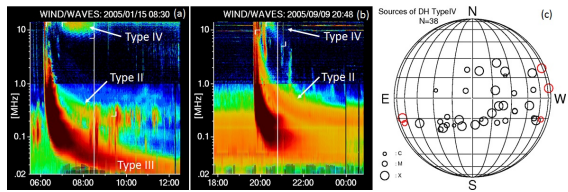

Fig. 4. Type IV bursts from disk-center (a) and limb (b) events. The vertical lines mark the end of the bursts. In (c), the solar sources of 38 cycle-23 type IV sources are plotted on a heliographic grid. The red data points indicate close-to-limb events (central meridian distance ≥ 60º). The size of the data points indicate the size of the associated soft X-ray flare.

IV. SUMMARY AND CONCLUSIONS

We compared the radio dynamic spectra of the 2013 November 7 event observed by 3 widely-spaced spacecraft and found that the type IV burst was only complete in STB, which was directly above the source region. The burst was partial in STA, which observed the burst and CME as a limb events. Wind did not see the burst at all. Taken together, these observations suggest that the type IV emission is directed along a narrow cone (less than ≈ 60º in width) from above the flare site. Thus the observer needs to be relatively close in angle to observe the complete burst. When the observer is at a large angle from the cone axis, the burst will appear partial. We also confirmed the directivity of type IV bursts statistically using all events from cycle 23: bursts originating from close to the disk center always appeared complete, while those from close-to-limb sources appeared partial.

The close connection between the flare site and the type IV burst suggests that the source of energy for the bursts is clearly the flare: electrons accelerated due to flare reconnection are trapped in the post-eruption structures (stationary type IV burst) producing radio emission at the local plasma frequency. The emission mechanism is likely to be coherent plasma process because Razin suppression is severe at low frequencies, so the gyro-synchrotron process may not be able to explain the bursts. The type IV bursts considered here are also likely to be stationary type IVs (similar to the flare continuum at metric wavelengths, see [10]) because moving type IV bursts are associated with moving CME structures. Moreover, these structures (CME flux ropes or prominences) have a larger angular extent than the post-eruption arcades, and hence should not show directivity. The extended nature of the CME shock is clear because the type II burst, including a 3-hour enhancement 7 hours after event onset, was observed in all three views from the highest frequencies (16 MHz in STEREO and 14 MHz in Wind). This is further confirmed by the observation of SEP events at all three spacecraft.

*Acknowledgment*

We thank the SOHO and STEREO teams for making the data freely available.